\documentclass{rspublic}

\begin{document}

\newcommand{\ea}{{\it et al. }}
\newcommand{\apj}{{\it Astrophys. J.}}
\newcommand{\aj}{{\it Astron. J.}}
\newcommand{\mnras}{{\it Mon. Not. R. Astron. Soc.}}
\newcommand{\aanda}{{\it Astron. Astrophys.}}
\newcommand{\ptra}{{\it Phil. Trans. R. Soc. A}}

\title[A revolution in star cluster research]{A revolution in star
cluster research:\\setting the scene}

\author[R. de Grijs]{Richard de Grijs}

\affiliation{Kavli Institute for Astronomy \& Astrophysics, Peking
University, Yi He Yuan Road 5, Haidian District, Beijing 100871,
People's Republic of China\\
Department of Physics \& Astronomy, The University of Sheffield, Hicks
Building, Hounsfield Road, Sheffield S3 7RH, UK}

\label{firstpage}

\maketitle

\begin{abstract}{{\bf ISM: kinematics and dynamics; globular clusters:
general; open clusters and associations: general; galaxies: evolution;
galaxies: star clusters; galaxies: stellar content}} Star clusters and
their stellar populations play a significant role in the context of
galaxy evolution, across space (from local to high redshift) and time
(from currently forming to fossil remnants). We are now within reach
of answering a number of fundamental questions that will have a
significant impact on our understanding of key open issues in
contemporary astrophysics, ranging from the formation, assembly and
evolution of galaxies to the details of the star-formation
process. Our improved understanding of the physics driving star
cluster formation and evolution has led to the emergence of crucial
new open questions that will most likely be tackled in a systematic
way in the next decade.
\end{abstract}

\section{Our improved understanding of star cluster physics}

It is now widely accepted that stars do not form in isolation, at
least for stellar masses above $\sim 0.5$ M$_\odot$. In fact, 70--90\%
of stars may form in a clustered mode (cf. Lada \& Lada 2003). Star
formation results from the fragmentation of molecular clouds, which in
turn preferentially leads to star cluster formation. Over time,
clusters dissolve or are destroyed by interactions with molecular
clouds or tidal stripping by the gravitational field of their host
galaxy. Their member stars become part of the general field stellar
population. Star clusters are thus among the basic building blocks of
galaxies. Star cluster populations, from young associations and open
clusters to old globular clusters, are therefore powerful tracers of
the formation, assembly and evolutionary history of their host
galaxies.

Using our improved understanding of star cluster physics, we are now
within reach of answering a number of fundamental questions in
contemporary astrophysics, ranging from the formation and evolution of
galaxies to the details of the process of star formation itself. These
two issues are the backbone of research in modern astrophysics. They
lead to new questions related to the make-up of the cluster and field
stellar populations in a variety of galaxies and galaxy types, their
relative formation timescales (and what this implies for overall and
resolved galactic star-formation histories) and the relationships
between local and high-$z$ stellar and cluster populations.

The refereed contributions in this issue of {\ptra} cover a wide range
of topics in contemporary astrophysics related to star cluster
formation, evolution, destruction and environmental impact,
particularly in the context of star formation on galactic scales. Lada
(2010) and Clarke (2010) discuss the formation of star clusters from,
respectively, an observer's and a simulator's point of view. Kalirai
\& Richer (2010) and van Loon (2010) then take detailed looks at
stellar and chemical evolution of star clusters and their constituent
stars, while Goodwin (2010) and Vesperini (2010) focus specifically on
aspects related to the binary stellar populations affecting star
cluster evolution and the makeup of the general field, and cluster
dynamics, respectively. Larsen (2010) and Harris (2010) take
wide-angle views of entire cluster populations, roughly split between
young (Larsen 2010) and old (Harris 2010) samples, while Bruzual
(2010) discusses their integrated evolution and issues relevant to the
assumptions usually adopted for modelling these systems as `simple
stellar populations' (SSPs).

While the review articles in this volume focus in detail on the
wide-ranging fields touched by state-of-the-art star cluster research,
here I address---non-exhaustively and roughly as a function of cluster
age---some of the higher-level challenges limiting sustained progress,
which will most likely be tackled by the research community in the
next decade.

\section{Early evolution}

One of the critical remaining issues pivotal to our understanding of
the early evolution of star clusters is the question as to precisely
how and when stars form. Although I will focus on the issues pertinent
to star cluster research, solving this issue will clearly have
profound consequences for a much wider range of fields in
astrophysics. Simplistically, the problem can be divided into two
subquestions: (i) How is star formation triggered and how does it
proceed, and (ii) How do the star-formation mode, efficiency and
pressure of the interstellar medium (ISM) lead to the resulting
stellar mass distribution (i.e., the initial mass function, IMF) and
in particular what is the role of the massive stars compared to that
of their lower-mass counterparts? The latter links the early evolution
of star clusters unequivocally to their environmental impact, which I
will address in \S3$\,a$.

\subsection{The low-mass stellar initial mass function}

Although the shape of the stellar IMF in the solar neighbourhood, and
particularly for stellar masses $> 1$ M$_\odot$, has essentially been
unchallenged since Salpeter's (1955) seminal study, its origin remains
hotly debated (e.g., Bonnell \ea 2007; Goodwin \& Kouwenhoven
2009). Constraining the physical origin of the IMF will have a major
impact on, e.g., our understanding of the conditions prevailing in a
wide range of starburst events, and the formation of the first stars
and clusters in the early universe, at $z\gtrsim5$---although, for a
full understanding, we would need to follow the radiative cooling
processes from primordial gas and the subsequently formed metallic
elements in full detail!

Significant uncertainties in shape remain at both the low- and
high-mass extremes of the IMF, however. At the low-mass end, the
prevailing models agree that the solar-neighbourhood IMF
flattens. This can be modelled by either multiple power-law or
lognormal mass distributions (cf. Kroupa 2001; Chabrier 2003). While
the former provides a mathematically simple and observationally useful
scaling law, the latter is supported by realistic numerical
simulations (Hennebelle \& Chabrier 2008). These take into account
dynamical depletion of the lowest-mass stars and replace the idea of a
single Jeans mass for all newly formed stars in a given molecular
cloud by a distribution of local Jeans masses which are representative
of the lognormal density distribution of the turbulent, fragmenting
gas. As statistically significant samples of roughly coeval stars,
rich young star clusters play a major role in constraining the
low-mass IMF. Open questions remaining in this field relate to whether
there is any metallicity dependence of the IMF shape for stellar
masses $< 1$ M$_\odot$, the initial structure of newly formed
clusters, and whether the ubiquitous mass segregation observed in
clusters of any age is dynamical or perhaps primordial (i.e., related
to the process of star formation).

Preliminary clues as to the shape of the low-mass IMF (down to $\sim
0.15$--0.30 M$_\odot$) in the low-metallicity ($Z \sim 0.4$ Z$_\odot$)
environment of young ($\sim 4$--45 Myr) Large Magellanic Cloud (LMC)
clusters have recently been uncovered on the basis of deep {\sl Hubble
Space Telescope (HST)} imaging observations (e.g., Da Rio \ea 2009;
Liu \ea 2009{\it a,b}). These studies imply that the IMFs of these
young clusters are essentially the same as that in the solar
neighbourhood, although the characteristic stellar masses are somewhat
higher. One would ideally want to probe younger star-forming regions
to reach firmer conclusions, but these are inevitably obscured by
large amounts of dust, hence requiring deep and often wide-field
infrared (IR), (sub)millimetre, radio and X-ray surveys (and pointed
observations) that are now coming online (e.g., the {\sl Spitzer Space
Telescope}'s GLIMPSE survey or the UKIRT IR deep-sky survey, UKIDSS;
e.g., Benjamin \ea 2003; Lucas \ea 2008) and which probe the low-mass
stellar mass distribution in particular (see, e.g., Rathborne \ea
2009).

\subsection{Initial substructure and mass segregation}

Simulations of star cluster evolution almost always assume that the
stars are initially smoothly distributed and in dynamical
equilibrium. However, both observations and the theory of star
formation tell us that this is not how clusters form. Goodwin \&
Whitworth (2004) and, more recently, Allison \ea (2009) investigated
the effects of substructure and initial clumpiness on the early
evolution of clusters. Comparisons with observations will allow us to
constrain how much initial substructure can be present. The most
massive stars in young star clusters are almost always observed to be
in the inner regions of those clusters (e.g., Hillenbrand \& Hartmann
1998; de Grijs \ea 2002{\it a,b,c}; Gouliermis \ea 2004). A crucial
question triggered by this observation relates to the origin of this
observed mass segregation. Do massive stars form in the centres of
clusters, or do they migrate there over time due to gravitational
interactions with other cluster members? In smooth, relaxed clusters
it seems that the most massive stars must form in the cores, which is
therefore often referred to as primordial mass segregation (but see
Ascenso \ea 2009). But does substructure play a role?

Both observational evidence (e.g., Larson 1995; Testi \ea 2000;
Elmegreen 2000; Lada \& Lada 2003; Gutermuth \ea 2005; Allen \ea 2007)
and theorical considerations suggest that young star clusters tend to
form with a significant amount of substructure. Their progenitor
molecular clouds are observed to have significant levels of
substructure in both density and kinematics (e.g., Carpenter \& Hodapp
2008), which is likely induced by the supersonic turbulence thought to
dominate molecular cloud structure (e.g., Mac Low \& Klessen 2004;
Ballesteros--Paredes \ea 2007). Observations also imply that young
clusters lose their substructure on timescales of $<2$~Myr (e.g.,
Cartwright \& Whitworth 2004; Schmeja \ea 2008). Simulations suggest
that the only way in which this could happen is if clusters are born
dynamically cool (Goodwin \ea 2004; Allison \ea 2009; Lada 2010). On
the basis of these arguments, Allison \ea (2009) recently performed an
ensemble of {\it N}-body simulations aimed at exploring the earliest
phases of cluster evolution. They find that cool, substructured
clusters appear to mass segregate dynamically for stellar masses down
to a few M$_\odot$ on timescales of a few Myr. This is reminiscent of
the observational status of the Orion Nebula Cluster (e.g., Bonnell \&
Davies 1998; Allison \ea 2009; Moeckel \& Bonnell 2009). More work is
required to systematically address the most likely initial conditions
for cluster formation leading to the observed configurations.

\subsection{Initial binarity}

Simulations of star clusters also often tend to neglect the presence
of binary stars. Observations of local star-forming regions lead us to
suspect that all, or nearly all, stars form in binary or triple
systems (Goodwin \& Kroupa 2005; Duch\^ene \ea 2007; Goodwin \ea
2007). Such systems significantly affect the dynamical evolution of
the cluster, yet the initial binary fractions in dense star clusters
are poorly known. Almost all studies of binarity have been limi\-ted
to nearby solar-metallicity populations (see Duch\^ene 1999 and
Duch\^ene \ea 2007 for reviews). However, it might be expected that
metallicity (e.g., through its effects on cooling and hence on the
opacity limit for fragmentation) will play a role in the fragmentation
of cores to produce binary systems (Bate 2005; Goodwin \ea 2007).

The binary fractions in more distant, massive clusters have not yet
been studied thoroughly, because of observational
limitations---although statistical colour--magnitude analysis based on
artificial-star tests offer a promising alternative (e.g., Romani \&
Weinberg 1991; Rubenstein \& Bailyn 1997; Bellazzini \ea 2002; Cool \&
Bolton 2002; Zhao \& Bailyn 2005; Davis \ea 2008). However, all
clusters thus far studied in this way are old stellar systems, in
which dynamical evolution is expected to have altered the initial
binary population significantly. Efforts have begun to address this
issue for the much more distant young populous clusters in the LMC
(e.g., Elson \ea 1998). Hu \ea (2009) estimate that the binary
fraction in NGC 1818 in the mass range between 1.3 and 1.6 M$_\odot$
is $\sim 0.35$ for systems with an approximately flat mass-ratio
distribution, $q$, for $q>0.4$. This is consistent with a {\it total}
binary fraction of F stars of 0.6 to unity. Do high binary fractions
affect mass segregation at early times or the relaxation of
substructure? Do they leave observational signatures? NGC 1818 is
several crossing times old, so that the binary population should have
been modified by dynamical interactions. In particular, soft (i.e.,
wide) binaries are expected to have been destroyed by this
age. Therefore, the high binary fraction found for F stars suggests
that these binaries are relatively `hard' and able to survive
dynamical encounters.

\subsection{The high-mass end of the stellar IMF}

Understanding the origin of mass segregation may also help distinguish
models of massive star formation. In particular, are the masses of the
most massive stars set by the mass of the core from which they form
(e.g., Krumholz \ea 2007), or by competitively accreting mass due to
being located at a favourable position in the cluster (e.g., Bonnell
\ea 1998; see also Krumholz \ea 2005, and Bonnell \& Bate 2006)?
Allison {\ea}'s (2009) result showing that dynamical mass segregation
can occur on a few crossing timescales suggests that massive stars
could form in relative isolation in large cores and mass segregate
later, possibly avoiding the need for competitive accretion as
dominant process to form the most massive stars in the centre of a
cluster.

The formation of the most massive stars in a given stellar population
is thus riddled with uncertainties. Claims (as well as counterclaims)
abound in the literature of top-heavy IMFs (i.e., containing too few
low-mass stars compared to the solar-neighbourhood IMF) among resolved
star clusters, both in and beyond the Local Group of galaxies (e.g.,
Smith \& Gallagher 2001 versus Bastian \ea 2008; Kim \ea 2006 versus
Klessen \ea 2007 versus Espinoza \ea 2009; Harayama \ea 2008; McKee \&
Tan 2002; Dabringhausen \ea 2009). The relevant open questions that
require firmer answers address why there are so few massive stars in
most `normal' stellar populations (is this simply a matter of
small-number statistics and hence stochasticity?) and, consequently,
how they form (cf. Zinnecker \& Yorke 2007) and what their impact is
on the cluster environment. The physical conditions in clusters
containing significant numbers of massive OB stars are of prime
interest for planetary scientists: although planet formation may be
qualitatively different in dense clusters compared to the more
quiescent field, is the formation of planetary systems from
protoplanetary discs and subsequent planet growth inhibited or
promoted near OB stars (e.g., Hollenbach \ea 2000; Throop \& Bally
2005)? Clearly, this question also applies to further star formation
near such massive stars.

\subsection{Timescales and triggering of star and cluster formation}

Following on from our rather rudimental understanding of the role of
massive stars during cluster formation, the latter is still far from
well understood. The two main competing scenarios require either a
rapid collapse in a single crossing time (e.g., Elmegreen 2000, 2007;
Heitsch \& Hartmann 2008) or slow formation over many dynamical times
(e.g., Tan \ea 2006). Proponents of the former scenario suggest that
this can be facilitated by flow-driven cloud formation of atomic gas
aided by gravitational compression, with supersonic turbulence (or
internal feedback) and star/protostar interactions playing only minor
roles. Arguments offered in favour of a slower formation timescale
include observational details such as age spreads in star-forming
clusters (for recent new insights, including multiple main sequences
in individual clusters, see the review by Piotto 2009) and the
momentum flux of molecular outflows, where turbulence does play an
important role. Kinematical studies based on precision astrometry and
radial velocities will hopefully soon enable us to refine theoretical
models of turbulent cloud collapse and draw firmer conclusions on the
timescales of star formation in a clustered mode and their subsequent
evolution.

This leads us to the question as to how exactly star formation in
clusters and associations is triggered, and how much of the molecular
gas is converted into stars (i.e., the star-formation efficiency,
SFE). On the smallest scales, the effects of the first generations of
massive stars in a particular star-forming region, in the form of
expanding H{\sc ii} regions and pre-supernova shocked and ionized OB
winds, trigger ongoing star formation by destroying their natal
molecular clouds (e.g., Joung \& Mac Low 2006; Elmegreen \& Palou\v{s}
2007). On galaxy-wide scales, both gravitational and feedback
processes seem important for the formation of the most massive
clusters. However, this may simply be the result of the hierarchical
star-formation process, where the densest regions have the highest
SFEs. If this were the full scenario, how then do quiescent spiral
galaxies characterized by low SFEs manage to form extremely massive
clusters that may eventually become counterparts to the ubiquitous old
globular clusters (see for a review de Grijs \& Parmentier 2007)? Are
there environments today that are conducive to the formation of
massive clusters that may eventually become old globular cluster
counterparts, or did the oldest star clusters in our Milky Way form in
an entirely different star-formation mode? Alternatively, how do dwarf
galaxies form extremely massive star clusters without external
triggers?

\subsection{The star-formation efficiency}

Does the formation of the highest-mass star clusters need an external
trigger, and hence does their age distribution fully reflect the
underlying galactic interaction history? How does the inferred SFE
compare to equivalent values elsewhere in the same or other galaxies
undergoing more quiescent star formation? Alternatively, do the
highest-mass clusters (with masses of, say, $>10^6$ M$_\odot$) form
differently from their lower-mass counterparts, for instance through
large-scale mergers of cluster complexes (e.g., Bastian \ea 2006;
Fellhauer \ea 2009) and how does this differ from the formation of
tidal dwarf galaxies (e.g., Bournaud \ea 2008)? On the other hand, is
their formation simply a consequence of the hierarchy of star
formation? If these young massive clusters are not very rare
exceptions, where then are the descendants of all those massive star
clusters that presumably formed earlier on in their host galaxies?

Hydrodynamical cluster-formation modelling shows that SFEs on the
order of 30\% or higher are required to form a massive cluster that is
long-term stable (e.g., Brown \ea 1995; Elmegreen \& Efremov 1997;
Bastian \& Goodwin 2006). This is at least an order of magnitude
higher than the SFEs in normal spiral and irregular galaxies, or in
dwarf galaxy starbursts (see, for a review, Anders \ea 2007). SFEs as
high as 30\% or more, however, are observed in global and nuclear
starbursts triggered by massive gas-rich mergers, such as NGC 7252,
and ultra-luminous IR galaxies (Fritze--v. Alvensleben \& Gerhard
1994; Gao \& Solomon 2004).

In addition, observations of a gap in cluster ages but not in that of
the field stellar population in the LMC suggest that cluster formation
there took place in stages of enhanced star formation only, possibly
related to close encounters of the LMC with the Milky Way and/or the
Small Magellanic Cloud (SMC). Similarly, star cluster formation in
both M51 and M82 is found to have been significantly enhanced during
their last close encounters with their neighbouring galaxies (e.g., de
Grijs \ea 2001; Bastian \ea 2005; Smith \ea 2007), once again,
precisely when the overall star formation and the SFE are expected to
be enhanced. In interacting galaxies, the frequency of molecular cloud
collisions is expected to increase strongly. This will considerably
enhance star formation. Moreover, molecular clouds get
shock-compressed by external pressure, grow denser and more massive,
and this can drive up the SFE very efficiently (e.g., Jog \& Solomon
1992; Barnes 2004). Jog \& Das (1992, 1996) showed that a relatively
small increase in the external ambient pressure to values 3--4 times
the internal pressure within the molecular clouds in an undisturbed
galaxy can drive SFEs up to 70--90\%. Does this mean that the
molecular cloud structure in galaxies characterized by enhanced SFEs
is somehow different from that in more quiescent galaxies? We probably
need to wait until the {\sl Atacama Large Millimeter/submillimeter
Array (ALMA)} comes online before we can even begin to address this
question in more distant interacting galaxies. It probably implies,
however, that star formation is more fundamentally governed by the
content of {\it high density} gas, not the {\it overall} gas content
(Solomon \ea 1992; Gao \& Solomon 2004). Hydrodynamical modelling of
galaxies and galaxy mergers will thus need to account for a
multi-phase ISM and include a careful description of phase
transitions, star formation and feedback processes.

Despite significant recent theoretical and observational progress, the
quantitative importance of triggering and the effects of varying SFEs
in cluster formation remain major challenges.

\section{Environmental impact}

\subsection{Feedback and star cluster survival}
\label{feedback.sec}

Star clusters, and particularly their most massive member stars,
ionize their natal H{\sc ii} regions, inflate wind-blown bubbles and
outflows, and eventually explode as supernovae. The latter, in turn,
chemically enrich the ISM, drive turbulence, and may trigger secondary
star and cluster formation and power superbubbles (e.g., Silich \ea
2007, 2009; W\"unsch \ea 2008) and `superwinds' (e.g., Westmoquette
\ea 2007, 2008, 2009; Law \ea 2009 for the Milky Way) into the haloes
of their host galaxies. Star cluster feedback processes are therefore
of fundamental importance for our understanding of the overall
energetics and evolution of galactic-disc stellar populations.

Young {\it massive} star clusters are of particular interest in this
context, as they contain some of the most massive main-sequence,
supergiant, hypergiant and Wolf--Rayet stars, which are associated
with strong winds and supernova remnants that profoundly affect the
surrounding ISM. Assessing the relationship of the most massive
clusters to the local ISM is critical to validate formation scenarios
of massive star clusters, and in particular whether IMF variations
might be expected as a function of ISM pressure differences (e.g.,
McKee \& Tan 2002). Recently, wide-field survey products in hitherto
inaccessible spectral domains (e.g., GLIMPSE in the mid-IR) have
revealed a significant population of previously unknown ionized
bubbles and supernova remnants, likely produced by thus far hidden
star clusters. Efforts are underway to better characterize the
star-forming landscape in the Galactic plane using these and other
surveys (e.g., Lucas \ea 2008; Minniti \ea 2009).

Cluster winds are as yet poorly understood because it is not possible
to treat directed outflows self-consistently (while full 3D radiative
transport computations are still beyond reach), but their importance
for chemical enrichment of the ISM is profound. The amount of
radiative cooling present within the cluster volume seems important
for a fuller understanding of their impact. It may be possible, under
the right conditions, for the wind to cool sufficiently within the
cluster to generate a second stellar generation. Could this perhaps
explain the secondary main sequences (or self-enrichment) observed in
some globular clusters (e.g., Bedin \ea 2004; Piotto 2009)? What
exactly is the interplay between cluster winds and the ISM and how
does this depend on the ISM pressure (cf. Westmoquette \ea 2007,
2008)? Can cluster outflows be inhibited if the ISM pressure is
sufficiently high?

New key questions are emerging rapidly: What are the survival chances
of young, embedded star clusters beyond the first $\sim 10$ Myr, in
view of the disruptive effects of these large-scale outflows (see also
\S4$a$)? What is the initial distribution of {\it gravitationally
bound} cluster masses (see for a review de Grijs \& Parmentier 2007)?
This is, of course, linked to the conditions (e.g., the interstellar
pressure and density) under which bound objects form, which traces
back to the issue of whether star clusters form the top of the
hierarchy of star formation. What is the SFE, and in particular the
{\it cluster}-formation efficiency, in regions of intersecting and
colliding (super)winds?

Finally, and more speculatively, do these feedback processes have any
bearing on the difference in dark matter content between compact star
clusters (no need for dark matter), the population of faint dwarf
spheroidal galaxies in the Local Group (dark-matter dominated; e.g.,
Swaters \ea 2003), the newly discovered extended clusters in M31
(Huxor \ea 2005), and the population of ultracompact dwarf galaxies
(e.g., Gregg \ea 2009)? In the diagnostic diagram showing absolute
$V$-band magnitude as a function of half-mass radius (e.g., Huxor \ea
2005), the region between the dark-matter dominated objects and that
occupied by `normal' clusters is increasingly being filled in,
particularly by the extended star cluster population in M31
(A. M. N. Ferguson 2009, personal communication). Is there a regime
between these objects where there is a sudden step change in dark
matter content, or is this change more gradual? Ongoing and planned
projects, such as the Pan-Andromeda Archaeological Survey (PAndAS;
McConnachie 2009), may soon shed light on these questions. Their
answers will have far-reaching implications for our understanding of
the cosmological evolution of dark matter haloes and their
star-forming cores (e.g., Bullock \& Johnston 2005; Johnston \ea
2008).

\subsection{Products of binary evolution}
\label{binaries.sec}

Many exotic objects observed in star clusters, such as blue stragglers
(BSs), cata\-clysmic variables and X-ray sources, as well as the
putative intermediate-mass black holes, are believed to be related
to binary systems. Cluster environments are particularly interesting
in the context of binary systems: they can be created, altered and
destroyed by interactions with their many nearby neighbours. Naively,
one would expect that the gravitational interactions of stars in
clusters will ultimately lead to a core collapse to an infinite mass
in a finite time. Indeed, post-core-collapse (PCC) clusters do exist,
but their cores have not collapsed in such a dramatic
manner. Something must therefore eventually stop this process, most
likely the formation of binaries in the cluster cores during the late
collapse stages. In fact, the formation of a few hard binaries could
entirely halt this core collapse, even in the most massive clusters.

X-ray observations, in particular, have provided circumstantial
evidence for significant binary populations in star clusters. A much
higher fraction of X-ray-bright sources in our Milky Way is associated
with globular clusters than would be expected from the field stellar
population. These are believed to be short-period binaries in which
mass is transferred from a main-sequence onto a neutron star
(cf. Phinney \& Kulkarni 1994). These low-mass X-ray binaries form
preferentially in globular clusters, perhaps as the result of core
collapse. However, many PCC clusters do not contain X-ray binaries, so
other types of binaries must exist there to halt their collapse.

Perhaps the most common binary collision products in clusters include
the ubiquitous BSs. Although BSs are relatively rare in number
compared to the regular member stars in a given cluster, these
luminous stars have a non-negligible effect on the cluster's
integrated-light properties. Nevertheless, their contributions to the
most commonly used SSP models are routinely neglected. Preliminary
results show that the integrated spectral properties of a sample of
Galactic open and rich Magellanic Cloud clusters are dramatically
modified by their BS components (Xin \& Deng 2005; Xin \ea 2007;
Y. Xin 2009, personal communication). Using either spectra or
broad-band colours, the resulting ages and/or metallicities will be
underestimated significantly. Conservatively, the underestimates to
both cluster ages and metallicities are $\sim50$\%. Given the ubiquity
of BSs in a great variety of environments, this seems an issue that
needs to be addressed rather urgently.

\section{Death throes}

Star clusters are subject to a variety of internal and external
mechanisms that, under the appropriate conditions, will
gravitationally unbind and subsequently disrupt them. These effects
include (see de Grijs \& Parmentier 2007), approximately as a function
of increasing timescale, (i) formation in a marginally bound state
(see also the review by Mac Low \& Klessen 2004), (ii) rapid removal
of the intracluster gas due to adiabatic or explosive expansion driven
by stellar winds or supernova activity, typically on timescales much
shorter than the proto-cluster dynamical crossing time, (iii) mass
loss due to normal stellar evolution (including the effects of stellar
winds and supernova explosions), (iv) internal two-body relaxation
effects, leading to dynamical mass segregation and the preferential
ejection of lower-mass stars, (v) release of energy stored in a
significant fraction of primordial hard binary systems, and (vi) tidal
and gravitational effects due to interactions with other significant
mass components, spiral arms, bulge or disc shocking and dynamical
friction.

\subsection{Cluster infant mortality}
\label{infant.sec}

Observations of increasing numbers of interacting and starburst
galaxies show a significantly larger number of young ($\lesssim
10$--30 Myr) star clusters than expected from a simple extrapolation
of the cluster numbers at older ages, taking into account the
observational completeness limits and the effects of sample binning,
and under the additional, simplifying assumption that the star cluster
formation rate has been roughly constant over the host galaxy's
history (see for reviews de Grijs \& Parmentier 2007; Whitmore \ea
2007).

These observations have prompted a flurry of activity in the area of
cluster disruption processes. This has led to suggestions that cluster
systems appear to be affected by a disruption mechanism that acts on
very short timescales ($\lesssim 10$--30 Myr) and which may be
mass-independent, at least for masses $\gtrsim 10^4$ M$_\odot$ (e.g.,
Fall \ea 2005; Bastian \ea 2005; Fall 2006). This fast disruption
mechanism, which is thought to effectively remove up to 50--90\% of
the youngest clusters from a given cluster population (e.g., Lada \&
Lada 1991; Whitmore 2004; Bastian \ea 2005; Mengel \ea 2005; Goodwin
\& Bastian 2006; Whitmore \ea 2007), is in essence caused by the rapid
removal of the intracluster gas on timescales of $\lesssim 30$
Myr. The observational effect resulting from this rapid gas removal
has been coined cluster `infant mortality' (Lada \& Lada 2003); it was
originally reported in the context of the number of very young
embedded clusters in the Milky Way, compared to their older, largely
gas-free counterparts.

The general consensus emerging from recent studies into these effects
is that rapid gas removal from young star clusters, which could leave
them severely out of virial equilibrium, would be conducive to
subsequent cluster disruption (Vesperini \& Zepf 2003; Bastian \ea
2005; Fall \ea 2005). The efficiency of this process will be enhanced
if a cluster's SFE is less than about 30\%, independent of the mass of
the cluster (see de Grijs \& Parmentier 2007 for a review). Goodwin \&
Bastian (2006) show that this type of cluster destruction occurs in
10--30 Myr (see also Kroupa \& Boily 2002; Lada \& Lada 2003; Lamers
\& Gieles 2008). The consequence of this is that clusters will expand
rapidly to attain a new virial equilibrium, and hence disappear below
the observational detection limit on a similar timescale. Depending on
their SFE, a fraction of the more tightly bound clusters will, by the
time they reach an age of $\sim 30$--40 Myr, subsequently contract
again (Goodwin \& Bastian 2006), hence increasing their mean surface
brightness and thus their chances of being detected in
magnitude-limited photometric surveys.

The early evolution of the star cluster population in the SMC has been
the subject of considerable recent interest and debate (e.g., Rafelski
\& Zaritsky 2005; Chandar \ea 2006; Chiosi \ea 2006; Gieles, Lamers \&
Portegies Zwart 2007; de Grijs \& Goodwin 2008). Chandar \ea (2006)
argued that the galaxy has been losing up to 90\% of its star clusters
per decade of age, at least for ages from $\sim 10^7$ up to $\sim
10^9$ yr, while Gieles \ea (2007) concluded that there is no such
evidence for a rapid decline in the cluster population, and that the
decreasing number of clusters with increasing age is simply caused by
evolutionary fading of their stellar populations in a
magnitude-limited cluster sample. de Grijs \& Goodwin (2008) set out
to shed light on this controversy (see also Lamers 2008). On the basis
of an independent data set, they placed a limit on the extent of
infant mortality between the age ranges $\sim3$--10 Myr to $\sim
40$--120 Myr of $\lesssim 30$\% ($1\sigma$). They ruled out a $\sim
90$\% mortality rate per decade of age at a $>6 \sigma$ level. In
addition, a first glance at the LMC cluster population's age
distribution (de Grijs \& Goodwin 2009) indicates that the number of
clusters populating the first $\sim 100$ Myr can likely also be fully
explained by simple evolutionary fading of a magnitude-limited cluster
sample, without the need to invoke infant mortality for cluster masses
$\gtrsim 10^3$ M$_\odot$.

These results raise a number of important questions. In particular,
could the apparent absence of infant mortality be hidden by
assumptions of a constant cluster formation rate? Alternatively (or
additionally), could mass-dependent infant mortality be at work in the
Magellanic Clouds? Or does the apparent difference between the
Magellanic Clouds on the one hand and the Antennae system, M51 and the
Milky Way on the other suggest a completely different underlying
physical process which may be density dependent? Finally, do the
effects of gas expulsion differ if we properly include realistic
initial conditions for star cluster formation in a combined
hydrodynamical/{\it N}-body modelling approach (cf. Fellhauer \ea
2009)?

\subsection{Dynamical dissolution}

Those clusters that survive the infant mortality phase will be subject
to the processes driving longer-term star cluster dissolution (see de
Grijs \& Parmentier 2007 for a review). The longer-term dynamical
evolution of star clusters is determined by a combination of internal
and external timescales. The free-fall and two-body relaxation
timescales, which depend explicitly on the initial cluster mass
density (e.g., Spitzer 1958; Chernoff \& Weinberg 1990; de la Fuente
Marcos 1997; Portegies Zwart \ea 2001), affect the cluster-internal
processes of star formation and mass redistribution through energy
equipartition, leading to mass segregation and, eventually, core
collapse. Internal relaxation will, over time, eject both high-mass
stars from the core (e.g., due to interactions with hard binaries (Can
we detect high-proper-motion escapers and predict their orbits and
observational properties, possibly using numerical approaches?) and
lose lower-mass stars from its halo through diffusion. However, the
external processes of tidal disruption, disc and bulge shocking, and
stripping by the surrounding galactic field (see, e.g., De Marchi \ea
2006) are in general more important for the discussion of this
disruption phase. Tidal disruption is enhanced by stellar evolution,
leading to mass loss through winds and/or supernova explosions, which
will further reduce the stellar density in a cluster, and thus make it
more sensitive to external tidal forces.

The remaining key open questions related to star cluster disruption
beyond the infant mortality phase appear mostly
observational. However, the more fundamental issues that might affect
our understanding of star cluster disruption relate to the accuracy of
the age determinations of the individual clusters, on which
statistical disruption analyses are based. Although contemporary
studies take account of the so-called `chimneys' in age space around
10 and 100 Myr (see, e.g., Bastian \ea 2005; de Grijs \& Goodwin
2008), caused by the onset of, respectively, red supergiants and
asymptotic giant-branch stars in normal stellar populations (neither
of which are as yet robustly implemented in any of the suites of SSP
models commonly used to convert multi-passband spectral energy
distributions into robust age estimates), the more fundamental
question to ask is what the realistic uncertainties are in the derived
age distributions.

In addition to the uncertain contributions of BSs (see \S3$b$), which
may artificially bias age estimates for intermediate-age and older
unresolved clusters, at younger ages ($\lesssim 30$ Myr) the key
uncertainty relates to the discrepancy of approximately a factor of
two between the pre-main-sequence contraction and nuclear
main-sequences age scales for resolved clusters. Naylor (2009)
suggests that by accounting for this discrepancy, the well-known lack
of clusters with ages in the range from 5 to 30 Myr (Jeffries \ea
2007) may disappear. In addition, and perhaps more speculatively, he
implies that this may also alleviate the problem that the planetary
disc-clearing timescale of $\sim 3$ Myr for young stars, as measured
from IR observations (e.g., Brice\~no \ea 2007) does not match the
$\sim 9$ Myr timescale required for planetary formation through
classical core accretion (cf. Pollack \ea 1996).

\section{Concluding thoughts}

Although the importance of understanding star cluster physics (e.g.,
in mapping out the Milky Way or as SSP probes) had been recognized for
decades, major progress has only become possible in recent years, both
for Galactic and extragalactic cluster populations. We have seen a
major recent investment in time and effort, largely thanks to
significant new resources in theory, simulations and
observations. These include breakthroughs in computational power (such
as the recent embrace of graphics processing units by the
computational astrophysics community), the maturing of {\sl
HST}-driven science (and the new generation of instruments as well as
upgrades and refurbishment of the older facility instruments), deep
and more precise (photometric, spectroscopic and astrometric) data for
large numbers of Galactic clusters spaning a significant age range
(including adaptive-optics-assisted imaging opportunities with the
largest ground-based telescopes, and wide-field developments
associated with intermediate-size apertures), and an explosion of
astrometric data.

In addition, rapid progress has been facilitated by the coming online
of observing facilities enabling access to hitherto unavailable
spectral regions, both on the ground and in space. The latter include,
among others, the {\sl Spitzer Space Telescope} and the Japanese {\sl
AKARI} satellite at IR wavelengths, the {\sl GALaxy Evolution eXplorer
(GALEX)} in the ultraviolet, the Japanese {\sl Suzaku} and the
European {\sl X-ray Multi-Mirror} mission {\sl (XMM--Newton)}, and the
{\sl Chandra X-ray Observatory} at shorter wavelengths. These
facilities will soon be complemented by the {\sl Herschel Space
Observatory} once commissioning has been completed, and in the
intermediate and long term by the {\sl James Webb Space Telescope} and
{\sl Gaia}, respectively. Ground-based precursors to the {\sl ALMA}
and upgrades of and extensions to existing (sub)millimetre and radio
facilities worldwide are starting to enable access to the full
submillimetre to metre-wave spectral range, with new facilities like
the {\sl Square Kilometer Array} due to come online in the next
decade. Recent advances in instrumentation are driving a renaissance
in the study of Galactic clusters (e.g., addressing the relationships
among open, globular and young massive clusters, and narrowing down
the systematic uncertainties hampering determinations of absolute
cluster ages in the Milky Way), while extragalactic cluster studies
are significantly aided by the development of new instrumentation
supporting ever wider fields of view.

Putting these results and new developments into the broader context of
galaxy evolution is the next logical step, which requires the combined
efforts of theorists, observers and modellers working on a large
variety of spatial scales, and spanning a very wide range of
expertise. I am confident that the next decade will see a continued
high level of research activity related to the physics driving star
cluster formation and evolution, with perhaps an enhanced focus on
their impact on the physics of their host galaxies. Given the exciting
new observational facilities due to come online in the next few years,
combined with major progress on computational and theoretical fronts,
a bright future no doubt awaits this field!

\begin{acknowledgements}
I thank Thijs Kouwenhoven and Simon Goodwin for carefully reading and
commenting on an early draft of this article. I also thank all authors
contributing to this volume for their enthusiastic support and the time
investment required to ensure the compilation of a high-quality
collection of review articles.
\end{acknowledgements}

\label{lastpage}
\end{document}